\documentclass[aps,pra,twocolumn]{revtex4}
\usepackage{amssymb}
\usepackage{epsfig}
\usepackage{bm}

\def\be{\begin{equation}}
\def\ee{\end{equation}}
\def\bea{\begin{eqnarray}}
\def\eea{\end{eqnarray}}

\begin{document}

\title{Evolution of collective $N$ atom states in single photon superradiance%
}
\author{Anatoly A. Svidzinsky and Marlan O. Scully}
\affiliation{{\small {Institute for Quantum Studies and Dept. of Physics, Texas A\&M
Univ., College Station TX 77843}}\\
Applied Physics and Materials Science Group, Engineering Quad, Princeton
University, Princeton NJ 08544 }
\date{\today }

\begin{abstract}
We present analytical solutions for the evolution of collective states of $N$
atoms. On the one hand is a (timed) Dicke state prepared by absorption of a
single photon and exhibiting superradiant decay. This is in strong contrast
to evolution of a symmetric Dicke state which is trapped for large atomic
clouds. We show that virtual processes yield a small effect on the evolution
of the rapidly decaying timed Dicke state. However, they change the long
time dynamics from exponential decay into a power-law behavior which can be
observed experimentally. For trapped states virtual processes are much more
important and provide new decay channels resulting in a slow decay of the
otherwise trapped state.
\end{abstract}

\maketitle

The long standing problem of collective spontaneous emission from $N$ atoms 
\cite{Dick54,Erns68} is a subject of growing recent attention \cite%
{Scul06,Scul07,Ebe06,Svid08a,Svid08b,Mana08,Frie08} and debate \cite{Svid08c}%
. Effects of virtual processes are of particular current interest \cite%
{Svid08b,Mana08,Frie08,Svid08c}.

Here we consider a system of two level ($a$ and $b$) atoms, $%
E_{a}-E_{b}=\hslash \omega $. Initially there are no photons and one of the
atoms is in the excited state $a$, but we don't know which one. That is the
system is prepared in a collective $N-$atom state. The question then is how
such a collective state evolves with time.

Atoms interact with common electromagnetic field and the interaction
Hamiltonian is given by%
\[
\hat{H}_{\text{int}}=\sum_{\mathbf{k}}\sum_{j=1}^{N}g_{k}(\hat{\sigma}%
_{j}e^{-i\omega t}+\hat{\sigma}_{j}^{\dag }e^{i\omega t}) 
\]%
\begin{equation}
\times \left( \hat{a}_{\mathbf{k}}^{\dag }e^{i\nu _{k}t-i\mathbf{k\cdot r}%
_{j}}+\hat{a}_{\mathbf{k}}e^{-i\nu _{k}t+i\mathbf{k\cdot r}_{j}}\right) ,
\label{h1}
\end{equation}%
where $\hat{\sigma}_{j}$ is the lowering operator for atom $j$, $\hat{a}_{%
\mathbf{k}}$ is the operator of photon with wave vector $\mathbf{k}$, $g_{k}$
is the atom-photon coupling constant and $\mathbf{r}_{j}$ is the radius
vector of the atom $j$. Evolution of the atomic system is described by the
state vector%
\begin{equation}
|\Psi \rangle =\sum_{j=1}^{N}\beta (t,\mathbf{r}_{j})|b_{1}b_{2}\ldots
a_{j}\ldots b_{N}\rangle  \label{EQ1}
\end{equation}%
where $|b_{1}b_{2}...a_{j}...b_{N}>$ is a Fock state in which atom $j$ is in
the excited state $a$ and all other atoms being in the ground state $b$. We
disregard polarization effects, that is treat photons as scalar and assume
that initial state evolves slowly compared to the time of photon flight
through the atomic cloud (the opposite limit has been studied in \cite%
{Svid08a}).

Decay of an initial state occurs via real and virtual processes in which a
virtual photon is emitted and then reabsorbed. In particular, due to
counter-rotating terms in Hamiltonian (\ref{h1}) virtual processes couple
the single-atom excited states with those in which two atoms are excited. If
all virtual processes are taken into account then for a dense atomic cloud
evolution of the system is described by an integral equation with an
exponential kernel \cite{Mana08,Svid08b} 
\begin{equation}
\frac{\partial \beta (t,\mathbf{r})}{\partial t}=i\gamma \frac{N}{V}\int d%
\mathbf{r}^{\prime }\frac{\exp (ik_{0}|\mathbf{r}-\mathbf{r}^{\prime }|)}{%
k_{0}|\mathbf{r}-\mathbf{r}^{\prime }|}\beta (t,\mathbf{r}^{\prime }),
\label{f1}
\end{equation}%
where $V=4\pi R^{3}/3$ is the volume of the spherical atomic cloud, $%
k_{0}=\omega /c$ and $\gamma $ is the single atom decay rate. We assume that
atoms are uniformly distributed with density $N/V$ in a sphere of radius $R$.

If we ignore virtual contributions then Eq. (\ref{f1}) {reduces to an
equation with sinusoidal kernel }%
\begin{equation}
\frac{\partial \beta (t,\mathbf{r})}{\partial t}=-\gamma \frac{N}{V}\int d%
\mathbf{r}^{\prime }\frac{\sin (k_{0}|\mathbf{r}-\mathbf{r}^{\prime }|)}{%
k_{0}|\mathbf{r}-\mathbf{r}^{\prime }|}\beta (t,\mathbf{r}^{\prime }).
\label{f1a}
\end{equation}%
Here we solve Eqs. (\ref{f1}) and (\ref{f1a}) analytically for two initial
conditions, namely the $|+>$ \textquotedblleft timed\textquotedblright\
Dicke state 
\begin{equation}
\beta (0,\mathbf{r})=e^{i\mathbf{k}_{0}\cdot \mathbf{r}},  \label{z1}
\end{equation}%
which is prepared by absorption of a single photon with wave vector $\mathbf{%
k}_{0}$ ($k_{0}=\omega /c$) \cite{Scul06,Scul07}, and the symmetric Dicke
state \cite{Dick54}%
\begin{equation}
\beta (0,\mathbf{r})=1.  \label{z2}
\end{equation}%
For a large atomic sample $R\gg \lambda $ ($\lambda =2\pi c/\omega $ is the
wavelength of the emitted photon) the $|+>$ state (\ref{z1}) is
superradiant, while (\ref{z2}) is a trapped state undergoing very slow
decay. As we show below, virtual processes yield a small (yet interesting)
effect on evolution of the rapidly decaying $|+>$ state. Such states decay
mainly via real Weisskopf-Wigner spontaneous emission processes. However,
virtual processes can substantially modify the dynamics of trapped states
and provide a main channel of decay.

\begin{figure}[h]
\bigskip 
\centerline{\epsfxsize=0.46\textwidth\epsfysize=0.4\textwidth
\epsfbox{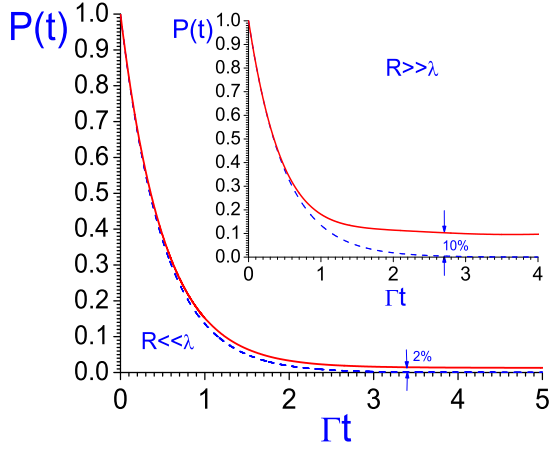}}
\caption{Probability that atoms are excited $P(t)$ for small and large
(insert) atomic clouds calculated using the $\exp $ (solid line) and $\sin $
(dash line) kernels. Initially atoms are prepared in the $|+>$ state, $%
\Gamma =N\protect\gamma $ (for $R\ll \protect\lambda$) and $\Gamma =3N%
\protect\gamma/2(k_0R)^2 $ (for $R\gg \protect\lambda $). For small atomic
sample the $|+>$ state and the symmetric state (\protect\ref{z2}) are the
same. }
\label{evop5}
\end{figure}

\begin{figure}[h]
\bigskip 
\centerline{\epsfxsize=0.45\textwidth\epsfysize=0.38\textwidth
\epsfbox{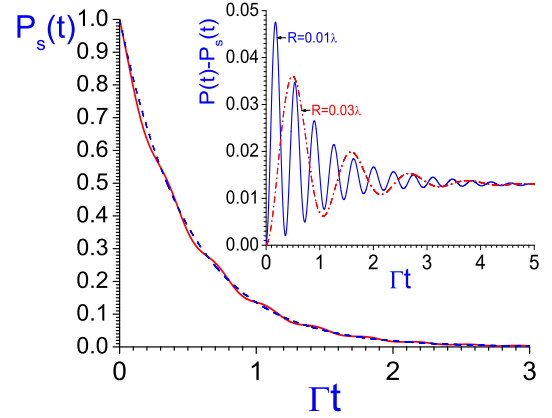}}
\caption{Probability that atoms are in the symmetric state $P_s(t)$ obtained
using the $\exp $ (solid line) and $\sin $ (dash line) kernels for $R=0.01%
\protect\lambda $. Initially atoms are in the symmetric state (\protect\ref%
{z2}) and $\Gamma =N\protect\gamma $. Insert shows probability to find atoms
in any other state but symmetric state calculated for $R=0.01\protect\lambda 
$ (solid line) and $R=0.03\protect\lambda $ (dash-dot line) from equation
with $\exp $ kernel.}
\label{evop2}
\end{figure}

\begin{figure}[h]
\bigskip 
\centerline{\epsfxsize=0.5\textwidth\epsfysize=0.4\textwidth
\epsfbox{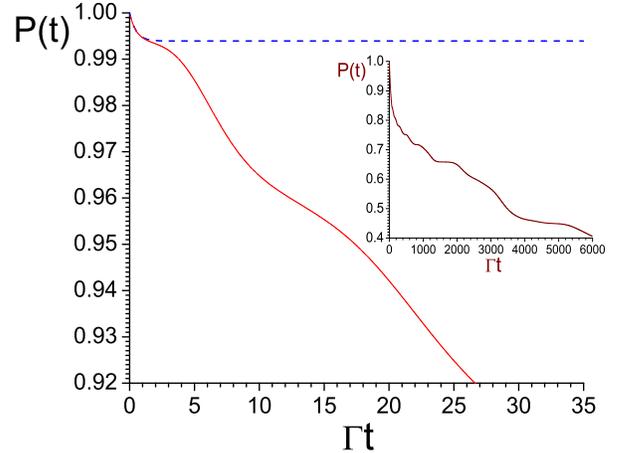}}
\caption{Probability that atoms are excited $P(t)$ obtained using the $\exp $
(solid line) and $\sin $ (dash line) kernels. Initially atoms are in the
symmetric state (\protect\ref{z2}), $R=5\protect\lambda$ and $\Gamma $ is
given by Eq. (\protect\ref{s2}).}
\label{unievop1}
\end{figure}

\begin{figure}[h]
\bigskip 
\centerline{\epsfxsize=0.45\textwidth\epsfysize=0.4\textwidth
\epsfbox{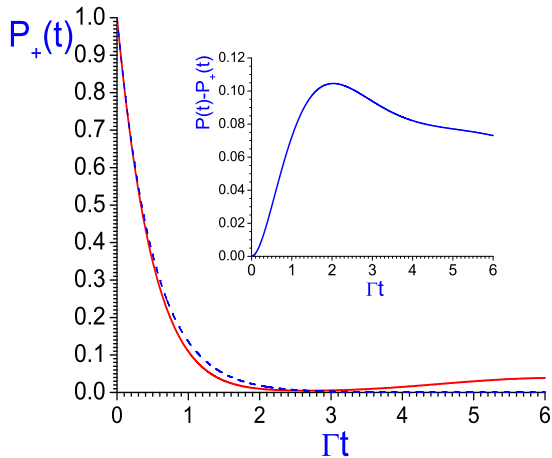}}
\caption{Probability that atoms are in the $|+>$ state $P_{+}(t)$ obtained
in the large sample limit from equation with $\exp $ (solid line) and $\sin $
(dash line) kernels. Insert shows probability that atoms are in any other
state but $|+>$. Initially atoms are in the $|+>$ state and $\Gamma $ is
given by Eq. (\protect\ref{s2}).}
\label{evop1}
\end{figure}

Figs. \ref{evop5}-\ref{evop1} summarize our main findings. For a small
atomic cloud $R\ll \lambda $ symmetric state (\ref{z2}) exponentially decays
according to Eq. (\ref{f1a}) with rate $\Gamma =N\gamma $ without coupling
to other states. This result has been obtained by Dicke \cite{Dick54}. Our
Figs. \ref{evop5} and \ref{evop2} show, however, that virtual processes
excite other states with a few $\%$ probability even in the small sample
(Dicke) limit. For a large cloud Eq. (\ref{f1a}) predicts that symmetric
state (\ref{z2}) is trapped, however, virtual processes lead to its slow
decay as shown in Fig. \ref{unievop1}. On the other hand, evolution of the
rapidly decaying $|+>$ state is only slightly affected by virtual processes
(see Figs. \ref{evop5} and \ref{evop1}). For a large sample such processes
excite other states with less then about $10\%$ probability. Thus the timed
Dicke state (\ref{z1}) is, to a good approximation, described by Eq. (\ref%
{f1a}) which ignores virtual transitions. However the symmetric state (\ref%
{z2}) is strongly effected by virtual processes as per Fig. \ref{unievop1}.

Next we discuss the evolution of $|+>$ state in detail. For $R\gg \lambda $
equation with $\sin $ kernel (\ref{f1a}) gives \cite{Scul07,Svid08b} 
\begin{equation}
\beta (t,\mathbf{r})=e^{i\mathbf{k}_{0}\cdot \mathbf{r}}e^{-\Gamma t},
\label{s1}
\end{equation}%
where 
\begin{equation}
\Gamma =\frac{3N\gamma }{2(k_{0}R)^{2}}.  \label{s2}
\end{equation}%
Here we find that equation with $\exp $ kernel (\ref{f1}) yields%
\begin{equation}
\beta (t,\mathbf{r})=e^{i\mathbf{k}_{0}\cdot \mathbf{r}}\left[
f(t,r)+ig(t,r)\cos \theta \right] ,  \label{e1}
\end{equation}%
where $\theta $ is the angle between $\mathbf{k}_{0}$ and $\mathbf{r}$, 
\begin{equation}
f(t,r)=\frac{1}{2}\left[ J_{0}\left( 2\sqrt{1-\frac{r}{R}}\sqrt{\Gamma t}%
\right) +J_{0}\left( 2\sqrt{1+\frac{r}{R}}\sqrt{\Gamma t}\right) \right] ,
\label{e2}
\end{equation}%
\begin{equation}
g(t,r)=\frac{1}{2}\left[ J_{0}\left( 2\sqrt{1-\frac{r}{R}}\sqrt{\Gamma t}%
\right) -J_{0}\left( 2\sqrt{1+\frac{r}{R}}\sqrt{\Gamma t}\right) \right]
\label{e3}
\end{equation}%
and $J_{0}(z)$ is the Bessel function. Answer (\ref{e1}) is universal in the
sense that state evolution is determined by the dimensionless time $\Gamma t$
and change of the sample size simply results in time rescaling.

To obtain solution (\ref{e1}) we used the identities 
\[
\frac{\exp (ik_{0}|\mathbf{r}-\mathbf{r}^{\prime }|)}{k_{0}|\mathbf{r}-%
\mathbf{r}^{\prime }|}=i\sum\limits_{m=0}^{\infty }(2m+1)P_{m}(\hat{r}\cdot 
\hat{r}^{\prime })\times 
\]%
\begin{equation}
\left\{ 
\begin{array}{c}
j_{m}(k_{0}r^{\prime })h_{m}^{(1)}(k_{0}r),\quad r>r^{\prime } \\ 
j_{m}(k_{0}r)h_{m}^{(1)}(k_{0}r^{\prime }),\quad r\leq r^{\prime }%
\end{array}%
\right. ,
\end{equation}%
\begin{equation}
\exp (i\mathbf{k}_{0}\cdot \mathbf{r})=\sum_{n=0}^{\infty
}i^{n}(2n+1)j_{n}(k_{0}r)P_{n}(\hat{k}_{0}\cdot \hat{r}),  \label{e4}
\end{equation}%
\begin{equation}
\int d\Omega _{r^{\prime }}P_{m}(\hat{r}\cdot \hat{r}^{\prime })P_{n}(\hat{k}%
_{0}\cdot \hat{r}^{\prime })=\delta _{mn}\frac{4\pi }{2n+1}P_{n}(\hat{k}%
_{0}\cdot \hat{r}),  \label{e8}
\end{equation}%
where $\hat{r}$ and $\hat{k}_{0}$ are unit vectors in the directions of $%
\mathbf{r}$ and $\mathbf{k}_{0}$ respectively, $P_{n}$ are the Legendre
polynomials and $j_{n}(z)$, $h_{n}^{(1)}(z)$ are the spherical Bessel
functions. In the large sample limit the ansatz (\ref{e1}) yields the
following equations for the slowly varying functions $f$ and $g$

\begin{equation}
\frac{\partial f(t,r)}{\partial t}=-\frac{\Gamma }{R}\int_{0}^{R}dr^{\prime
}f(t,r^{\prime })-\frac{\Gamma }{R}\int_{r}^{R}dr^{\prime }g(t,r^{\prime }),
\label{e11}
\end{equation}%
\begin{equation}
\frac{\partial g(t,r)}{\partial t}=\frac{\Gamma }{R}\int_{0}^{r}dr^{\prime
}f(t,r^{\prime }),  \label{e12}
\end{equation}%
with the initial conditions $f(0,r)=1$ and $g(0,r)=0$. One can solve Eqs. (%
\ref{e11}) and (\ref{e12}) using the method of Laplace transform which
yields the answer (\ref{e2}) and (\ref{e3}).

Next we calculate the probability that atoms are excited as a function of
time%
\begin{equation}
P(t)=\frac{1}{V}\int d\mathbf{r}|\beta (t,\mathbf{r})|^{2}.  \label{e17}
\end{equation}%
For the integral equation with $\sin $ kernel%
\begin{equation}
P_{\sin }(t)=e^{-2\Gamma t}.  \label{e18}
\end{equation}%
For $\beta (t,\mathbf{r})$ given by Eq. (\ref{e1}) one can calculate the
integral in Eq. (\ref{e17}) numerically for any $t$, while for $t\gg
1/\Gamma $ we find%
\begin{equation}
P_{\exp }(t)\approx \frac{7\sqrt{2}}{15\pi \sqrt{\Gamma t}}=\frac{0.21}{%
\sqrt{\Gamma t}}.  \label{e20a}
\end{equation}%
Insert of Fig. \ref{evop5} shows $P(t)$ obtained using the $\exp $ kernel
(solid line) and Eq. (\ref{e18}) (dash line). \ At $t\lesssim 1/\Gamma $ the
function $P_{\exp }(t)$ decays as $e^{-2\Gamma t}$, while for $t>1/\Gamma $
it becomes closer to its asymptotic expression (\ref{e20a}). During the
major part of the decay curve $P_{\exp }(t)$ exhibits exponential behavior (%
\ref{e18}) and, thus, virtual processes have essentially no effect. However,
virtual processes modify $P_{\exp }(t)$ at large time yielding the power-law
decay (\ref{e20a}). Such an interesting, although small, effect can be
observed experimentally.

For solution (\ref{e1}) the probability that atoms are in the $|+>$ state is
given by%
\[
P_{\exp }^{+}(t)=\frac{9}{4}\left( \frac{4(\Gamma t-2)}{(\Gamma t)^{2}}%
J_{0}(2\sqrt{2\Gamma t})+\right. 
\]%
\begin{equation}
\left. \frac{\sqrt{2}}{(\Gamma t)^{5/2}}[4-6\Gamma t+(\Gamma t)^{2}]J_{1}(2%
\sqrt{2\Gamma t})\right) ^{2}  \label{e21a}
\end{equation}%
which for $t\gg 1/\Gamma $ yields 
\begin{equation}
P_{\exp }^{+}(t)\approx \frac{9\sqrt{2}}{4\pi (\Gamma t)^{3/2}}\cos
^{2}\left( 2\sqrt{2\Gamma t}+\frac{\pi }{4}\right) .
\end{equation}

In Fig. \ref{evop1} we plot $P_{\exp }^{+}(t)$ obtained from Eq. (\ref{e21a}%
) (solid line) and compare it with those found from equation with $\sin $
kernel $P_{\sin }^{+}(t)=e^{-2\Gamma t}$ (dash line). The two curves are
very close to each other. This means that virtual processes practically do
not change evolution of $|+>$ state if it is considered separately. Without
virtual processes the $|+>$ state directly decays into the ground state by
emitting a photon. Virtual processes yield an extra decay channel in which
energy is partially transferred into other atomic states. However, as one
can see from Fig. \ref{evop1}, the net decay rate of the $|+>$ state into
all channels remains practically the same with or without virtual processes.
Insert shows probability that atoms are in any other state but $|+>$. This
curve demonstrates that during the system evolution the other states are
excited with probability less then about $10\%$ and, therefore, the effect
of virtual processes is quite small for fast decaying states.

Next we discuss evolution of the symmetric state (\ref{z2}). For such
initial condition Eq. (\ref{f1a}) with $\sin $ kernel can be solved
analytically for any size of the atomic sample and yields 
\begin{equation}
\beta (t,\mathbf{r})=1+2F\frac{\sin (k_{0}r)}{k_{0}r}\left[ 1-e^{-\Gamma t}%
\right] ,  \label{k7}
\end{equation}%
where%
\begin{equation}
F=\frac{k_{0}R\cos (k_{0}R)-\sin (k_{0}R)}{k_{0}R-\sin (k_{0}R)\cos (k_{0}R)}
\end{equation}%
and 
\begin{equation}
\Gamma =\frac{3\gamma N}{2(k_{0}R)^{2}}\left[ 1-\frac{\sin (2k_{0}R)}{2k_{0}R%
}\right] .  \label{k8}
\end{equation}%
Eq. (\ref{k7}) shows that at the beginning the atomic system decays with the
superradiant rate (\ref{k8}) but quickly ends up in a trapped state 
\begin{equation}
\beta (\mathbf{r})=1+2F\frac{\sin (k_{0}r)}{k_{0}r}.  \label{k9}
\end{equation}%
Function (\ref{k9}) vanishes in the small sample limit $k_{0}R\ll 1$,
however, for large sample $\beta (\mathbf{r})\approx 1$ and state (\ref{z2})
is completely trapped. Probability that atoms are excited is given by 
\begin{equation}
P(t)=1-\frac{6\left[ k_{0}R\cos (k_{0}R)-\sin (k_{0}R)\right] ^{2}}{%
k_{0}R-\sin (k_{0}R)\cos (k_{0}R)}\frac{\left[ 1-e^{-2\Gamma t}\right] }{%
(k_{0}R)^{3}}.  \label{z5}
\end{equation}%
For a large atomic cloud $R\gg \lambda $ the evolution Eq. (\ref{f1}) with
initial condition (\ref{z2}) can be also solved analytically and the answer
is expressed in terms of the Bessel functions. In Fig. \ref{unievop1} we
plot probability that atoms are excited $P(t)$ obtained from equation with $%
\exp $ (solid line) and $\sin $ (dash line) kernels. Initially atoms are
prepared in the state (\ref{z2}). Size of the atomic sample is $R=5\lambda $%
. Insert shows behavior of $P(t)$ for $\exp $ kernel on a large time scale
which exhibits interesting plateaus and oscillations. For $t$ less then a
few $1/\Gamma $ two curves are identical. For such time the real processes
dominate and the initial state evolves into the state (\ref{k9}) which is
trapped if we omit virtual processes. Virtual processes, however, result in
state decay as shown by the solid curve. State (\ref{z2}) overlaps with many
eigenstates of Eq. (\ref{f1}) \ \cite{Svid08b}. Eigenstates which decay
faster contribute to evolution at small time. As time increases $P(t)$
decays more slowly. However, eigenfunctions of Eq. (\ref{f1}) are not
orthogonal and, in addition, have different collective Lamb shifts. This
makes state evolution richer.

In the small sample limit $R\ll \lambda $ the initial states (\ref{z1}) and (%
\ref{z2}) are the same. Equation with $\sin $ kernel (\ref{f1a}) gives Dicke
result%
\begin{equation}
\beta (t,\mathbf{r})=e^{-\Gamma t},
\end{equation}%
where $\Gamma =N\gamma $. For equation with $\exp $ kernel the state
evolution can be obtained by noting that in the small sample limit 
\begin{equation}
\beta _{n}(t,r)=\frac{R}{r}\sin \left[ \left( \pi n+\frac{\pi }{2}\right) 
\frac{r}{R}\right] e^{-\lambda _{n}t}
\end{equation}%
are eigenfunctions of Eq. (\ref{f1}) with eigenvalues \cite{Svid08b} 
\begin{equation}
\lambda _{n}=-\frac{12iN\gamma }{\pi ^{2}(2n+1)^{2}k_{0}R}+\frac{96N\gamma }{%
\pi ^{4}(2n+1)^{4}},\quad n=0,1,2,\ldots
\end{equation}%
Using the identity 
\[
1=\frac{4}{\pi x}\sum_{n=0}^{\infty }\frac{(-1)^{n}}{(2n+1)^{2}}\sin
[(2n+1)x] 
\]%
one can expand the initial condition $\beta (0,r)=1$ in terms of $\beta
_{n}(0,r)$. As a result, time evolution of the symmetric state is given by%
\begin{equation}
\beta (t,\mathbf{r})=\frac{8R}{\pi ^{2}r}\sum_{n=0}^{\infty }\frac{(-1)^{n}}{%
(2n+1)^{2}}\sin \left[ \left( n+\frac{1}{2}\right) \frac{\pi r}{R}\right]
e^{-\lambda _{n}t}  \label{e32}
\end{equation}%
and probability to find atoms excited is%
\begin{equation}
P(t)=\frac{96}{\pi ^{4}}\sum_{n=0}^{\infty }\frac{\exp \left[ -2\text{Re}%
(\lambda _{n})t\right] }{(2n+1)^{4}}.  \label{e33}
\end{equation}

Fig. \ref{evop5} shows $P(t)$ given by Eq. (\ref{e33}) (solid line) and
compares it with the answer obtained omitting virtual processes $P(t)=\exp
(-2\Gamma t)$ (dash line). The two curves are close to each other, but Eq. (%
\ref{e33}) yields a few $\%$ of population trapped which slowly decays with
time.

In Fig. \ref{evop2} we plot probability that atoms are in the symmetric sate
(\ref{z2}) obtained for $R=0.01\lambda $ from Eq. (\ref{e32}) (solid line)
and compare it with $P(t)=\exp (-2\Gamma t)$ (dash line). The two curves are
very close meaning that the net decay rate of the symmetric state into all
channels is the same with or without virtual processes. Insert shows
probability to find atoms in any other state but symmetric sate (\ref{z2})
for $R=0.01\lambda $ (solid line) and $R=0.03\lambda $ (dash-dot line)
obtained from Eq. (\ref{e32}). Dependence of the imaginary part of $\lambda
_{n}$ (collective Lamb shift) on $n$ is the reason for oscillations. Period
of oscillations is proportional to $k_{0}R$. The other states are excited
with a few $\%$ probability. Thus, in the small sample limit, virtual
photons also yield a small (but interesting) effect on evolution of fast
decaying states.

In summary, we consider evolution of two collective states of $N$ atoms, the 
$|+>$ state which decays fast and the symmetric state which is trapped for $%
R\gg \lambda $. We obtain analytical formulas for the atomic state vector as
a function of time. We show that virtual processes yield a small effect on
evolution of the rapidly decaying states. However, they change the long time
dynamics from exponential decay into power-law which can be observed
experimentally. For trapped states virtual processes qualitatively modify
state evolution. Namely, they provide new decay channels which ultimately
result in a slow decay of the otherwise trapped state.

We thank R. Friedberg and J. Manassah for stimulating discussion and
gratefully acknowledge the support of the Office of Naval Research (Award
No. N00014-07-1-1084 and N0001408-1-0948) and the Robert A. Welch Foundation
(Award A-1261).

\end{document}